\documentclass[12pt]{article}
\def\IFUM{695/FT}
\usepackage{amsfonts}
\usepackage{a4}
\usepackage{amssymb}
\usepackage{amsmath}
\def\Red{}
\def\Black{}
\def\Blue{}
\def\back{\!\!\!\!\!\!\!\!\!}
\newcommand{\staranticomm}[2]{\{ #1 \, ,^{\!\!\star} #2 \}} 
\newcommand{\starcomm}[2]{[ #1 \, ,^{\!\!\star} #2 ]}
\newcommand{\stargr}[1]{\star^{\!\!\!^#1}}
\begin{document}     

\par

\begin{flushright}
{\bf IFUM \IFUM} \\
\end{flushright}
\vskip 0.2 truecm
{\Red
{\Large
{\bf
\begin{center}
Chern-Simons in the Seiberg-Witten map
for non-commutative Abelian gauge theories in $4D$
\end{center}
}
}
\vspace{0.2cm}
\Black}

\begin{center}
{\large
 Marco Picariello$^\mathrm{(a)}$\footnote{Marco.Picariello@mi.infn.it}, 
 Andrea Quadri$^\mathrm{(a)}$\footnote{Andrea.Quadri@mi.infn.it},
 Silvio P.~Sorella$^\mathrm{(b)}$\footnote{sorella@uerj.br}
\\
{\small
\vskip 0.2 truecm 
$\mathrm{(a)}$
Dipartimento di Fisica, Universit\`a di Milano
via Celoria 16, 20133 Milano, Italy \\
and  INFN, Sezione di Milano \\

\vskip 0.2 truecm
$\mathrm{(b)}$
UERJ, Universidade do Estado do Rio de
Janeiro\\
 Rua S\~{a}o Francisco Xavier 524, 20550-013 Maracan\~{a},
Rio de Janeiro, Brazil.
}
}
\end{center}

\vskip 0.2 cm

{\Blue
\centerline{\bf Abstract}
\begin{quotation}
\small
A cohomological BRST characterization of the Seiberg-Witten (SW) map is given.
We prove that the coefficients of the SW map can be identified with elements
 of the cohomology of the BRST operator modulo a total derivative.
As an example, it will be illustrated how the first coefficients of the SW
 map can be written in terms of the Chern-Simons three form.
This suggests a deep topological and geometrical origin of the SW map.
The existence of the map for both Abelian and non-Abelian case is discussed. 
By using a recursive argument and the associativity of the
 $\star$-product, we shall be able to prove that the Wess-Zumino consistency
condition for non-commutative BRST transformations is fulfilled.
The recipe of obtaining an explicit solution by use of the homotopy
 operator is briefly reviewed in the Abelian case.

\end{quotation}
\Black}

\newpage

\section{Introduction}

The construction of non-commutative gauge theories via the Seiberg-Witten (SW)
 map~\cite{Seiberg:1999vs} has been recently considered  by many
 authors~\cite{Asakawa:1999cu,Okuyama:2001sw,Jurco:2001rq,Brace:2001fj,Barnich:2001mc}.
The main issue is whether it is possible to
 deform~\cite{Jurco:2001rq,Brace:2001fj} the structure equations of ordinary
 gauge theories into the non-commutative counterpart. 

The commutative ghost and gauge fields $c$, $a_i$ take values in the Lie
 algebra $\mathfrak{G}$ and transform as 
\begin{subequations}\label{e1}
\begin{eqnarray}
&& s c = i c^2 \\
&& s a_i = \partial_i c + i [c, a_i] \,,
\end{eqnarray} 
\end{subequations}
where $s$ is the BRST differential. Analogously, $C$, $A_i$ denote the
 corresponding non-commutative fields, which are formal power series in 
 $\theta$ whose coefficients are local polynomials in  $c$ and $a_i$.
They are required to obey the conditions~\cite{Brace:2001fj} 
\begin{subequations}\label{e2}
\begin{eqnarray}
&& s C = i C \star C \\
&& s A_i = \partial_i C + i \starcomm{C}{A_i} \, .\label{e2:3}
\end{eqnarray}
\end{subequations}
The fields $C,A_i$ reduce to $c,a_i$ at $\theta=0$
\begin{subequations}\label{e4}
\begin{eqnarray}
&& C = c + O(\theta) \, , \label{e4:1}\\
&& A_i = a_i + O(\theta) \, .
\end{eqnarray}
\end{subequations}
In addition, $C$ is taken to be linear in $c$.
Moreover, it has been underlined~\cite{Jurco:2001rq} that $C, A_i$ 
 cannot be Lie-algebra valued if they have to fulfill eqs.(\ref{e2}).
They are  in fact elements of the enveloping algebra
 $Env(\mathfrak{G})$~\cite{Jurco:2001rq}.

For the star operation  we adopt the Weyl-Moyal product 
\begin{eqnarray}
f \star g & = &
\left . \exp \left( {i \over 2} \theta^{ij} {\partial \over \partial x^i}
{\partial \over \partial y^j} \right) f(x) g(y) \right |_{y \rightarrow x} 
\nonumber \\
& = & f(x)g(x) +  {i \over 2} \theta^{ij} \partial_i f(x) \partial_j g(x)
 + \dots  
\label{e3}
\end{eqnarray}
and 
\begin{eqnarray}
\starcomm{A}{B} &=& A \star B - B \star A \nonumber\\
		&=& i \theta^{ij} \partial_i A \partial_j B + \dots \, .
\label{e3.bis}
\end{eqnarray}
where $\theta^{ij}$ is an anti-symmetric constant matrix.

The solutions of eqs.(\ref{e2}) turn out to be affected by several
 ambiguities~\cite{Asakawa:1999cu,Jurco:2001rq,Brace:2001fj}, which can be
 conveniently analysed from a cohomological point of view.  Indeed, as shown
 in~\cite{Brace:2001fj}, a  nilpotent differential $\Delta$ is naturally
 generated when one tries to solve eqs.(\ref{e2}) order by order in the
 $\theta$ expansion.
In the Abelian case $\Delta$ reduces to the ordinary Abelian BRST differential.
Thus, the problem of finding an explicit solution of eqs.(\ref{e2}) is
 equivalent to the  construction of  a suitable homotopy operator for $\Delta$,
 provided that the corresponding Wess-Zumino consistency condition is
 fulfilled.
This is the approach taken in ref.\cite{Brace:2001fj}.
Recently, the homotopy operator relevant for the Freedman-Townsend Abelian
 non-commutative model has been obtained in~\cite{Barnich:2001mc}. 

The aim of the present work is to pursue the investigation of the cohomological
 properties of the SW map. We first focus on the Abelian case.
 Here we point out that the most general solution of eqs.(\ref{e2}) can be
 characterized in terms of elements of the cohomology of $s$ modulo $d$,
 $H(s|d)$. The aforementioned ambiguities show up as elements of the local
 cohomology of $s$, $H(s)$, and can be interpreted as field
 redefinitions, according to~\cite{Asakawa:1999cu,Jurco:2001rq,Brace:2001fj}.

It is worth reminding that the elements of the cohomology of $s$ modulo $d$
 can be written in terms of topological quantities like the Chern-Simons
 three-form and its generalizations.
As a consequence, we shall prove that the lowest order coefficients of the
 expansion of $C$ and $A_i$ can be  expressed indeed in terms of the
 Chern-Simons form, up to a total derivative.
This observation leads us to suggest  that the coefficients of
 the SW map possess a deep topological origin. 
  
We also check that the Wess-Zumino consistency condition of 
 ref.~\cite{Brace:2001fj} associated to eqs.(\ref{e2})
 is  satisfied due to  the associativity of the $\star$-product,
 both for the Abelian and the non-Abelian case.
This is a necessary condition for the existence of the SW map.
It becomes a sufficient condition if the homotopy operator for $\Delta$ is
 obtained. 
This point will be illustrated in detail for the Abelian case by providing an
 explicit formula for the homotopy.

The paper is organized as follows. In Sect.~\ref{2} we present the most general
 solution for the non-commutative transformations of the ghost and gauge field,
 and we shall check that the lowest order coefficients of the SW map can be
 cast in the form of the Chern-Simons.
Sect.~\ref{sec:WZ} is devoted to the recursive proof of the fulfillment of the
 Wess-Zumino consistency condition and of the existence of the SW map in the
 Abelian case.

\section{General solution of non-commutative Abelian gauge transformations}
\label{2}

In the Abelian case the BRST differential $s$ is 
\begin{eqnarray}
s a_i = \partial_i c \, , ~~~~ s c =0  \,.
\label{e5}
\end{eqnarray}
Besides the $\theta$-expansion, in this  case there is an 
 additional grading~\cite{Okuyama:2001sw}   
 which is useful in order to solve the equations
\begin{subequations}\label{e6}
\begin{eqnarray}
&& s C = i C \star C \, , \label{e6.1}\\
&& s A_i = \partial_i C + i \starcomm{C}{A_i} \, .\label{e6.2}
\end{eqnarray}
\end{subequations}
This grading consists in counting the number of gauge fields  $a_j$ present
 in a given expression.
This amounts to expand both $C$ and $A_i$ as 
\begin{subequations}\label{7}
\begin{eqnarray}
&& C = c + \sum_{n=1}^\infty C^{(n)}(a_j,c,\theta) \, , \\
&& A_i = a_i + \sum_{n=2}^\infty A^{(n)}_i(a_j,\theta) \, ,
\label{e7}
\end{eqnarray}
\end{subequations}
where $C^{(n)}$, $A_i^{(n)}$ are assumed to be of order $n$ in $a_j$.
Each coefficient $C^{(n)}$, $A_i^{(n)}$ can be further expanded 
 in powers of $\theta$.
As it can be seen from eq.(\ref{e5}), the BRST differential $s$ has degree
 $-1$ with respect to the degree induced by counting the fields $a_j$.

Let us begin with eq.(\ref{e6.1}). Expanding in powers of $a_j$ one gets 
\begin{eqnarray}
&& s c = 0 \, , \nonumber \\
&& s C^{(1)} = {i \over 2} \staranticomm{c}{c} \,  , \nonumber \\
&& s C^{(2)} = i \staranticomm{c}{C^{(1)}} \, , \nonumber \\
&& s C^{(3)} = i \staranticomm{c}{C^{(2)}} + 
{i \over 2} \staranticomm{C^{(1)}}{C^{(1)}} \, , \nonumber \\
&& \dots 
\label{e8} 
\end{eqnarray}
A particular solution of these equations has been found in~\cite{Jurco:2001rq}
 and takes the form
\begin{subequations}\label{e9}
\begin{eqnarray}
&& C^{(1)} = {1 \over 2} \theta^{ij} a_j \partial_i c \, , \label{e9:a} \\
&& C^{(2)} = {1 \over 6} \theta^{kl}\theta^{ij}
a_l ( \partial_k ( a_j \partial_i c) - f_{jk} \partial_i c ) \, ,
\label{e9:b}
\end{eqnarray}
\end{subequations}
where the expansion in $\theta$ has been truncated at the first non trivial
 orders and  $f_{jk}$ stands for the Abelian commutative field  strength 
\begin{eqnarray}
f_{jk} \equiv \partial_j a_k - \partial_k a_j \, .
\label{e10}
\end{eqnarray}
It should be observed that these expressions  provide only a particular
 solution  to eq.(\ref{e6.1}).
Other solutions have been found in~\cite{Okuyama:2001sw,Jurco:2001rq} and
 differ by elements which are BRST invariant. In fact, from the eqs.(\ref{e8}) 
 it follows that the coefficients $C^{(n)}$ are always defined modulo terms
 which are BRST invariant.
Moreover, as remarked in~\cite{Jurco:2001rq,Bichl:2001cq}, these
 extra terms should be taken into account in order to obtain a consistent
 quantum version of the non-commutative theories.

It is the purpose of the next subsection to give a precise
 cohomological interpretation of the general solution of eq.(\ref{e6.1}).

\subsection{The general solution for $C$}

Let us work out the most general solution of eq.(\ref{e6.1}). 
This equation  can be solved if and only if there exists a $\Lambda$ such that
\begin{eqnarray}
 i C \star C = {i \over 2} \staranticomm{C}{C} = s \Lambda  \, ,
\label{e11}
\end{eqnarray}
where $\Lambda$ is a local formal power series in  $a_j,\theta$ and linear
 in $c$.
If $\Lambda$ exists\footnote{We refer the reader to Sect.~\ref{sec:WZ}
 for a proof of the existence of $\Lambda$.}, $C$ must satisfy the homogeneous
 equation
\begin{eqnarray}
s ( \Lambda - C) = 0 \, ,
\label{e12}
\end{eqnarray}
which in turn implies
\begin{eqnarray}
C = \Lambda + \xi \, , ~~~~ \xi \in H(s) \, .
\label{e13}
\end{eqnarray}
In the above equation $\xi$ stands for a representative of $H(s)$, the
 local cohomology of $s$. Of course, $H(s)$ contains an infinite number of
 elements, as for instance  
\begin{eqnarray}
\theta^{ij} c f_{ij}\, , ~~~
\theta^{ij} \theta^{mn} c f_{ij} f_{mn} \, , ~~~
\theta^{ij} \theta^{mn} \theta^{pq} c f_{ij} f_{mn} f_{pq}  \, , ~~~\dots
\label{e14}
\end{eqnarray}
We can now proceed to characterize $\Lambda$.
For that purpose we recall a few properties of the $\star$ product.
From the Taylor expansion
\begin{eqnarray}
&&\back
f(x) \star g(x)  =  \left . \exp \left( {i \over 2} \theta^{ij} {\partial \over \partial x^i}
{\partial \over \partial y^j} \right) f(x) g(y) \right |_{y \rightarrow x} 
\nonumber \\
&& =  f(x)g(x) + \sum_{n=1}^\infty
{1 \over n!}  \left ( {i \over 2} \right )^n \theta^{i_1 j_1} \dots
\theta^{i_n j_n} \partial_{i_1} \dots \partial_{i_n} f(x) 
                 \partial_{j_1} \dots \partial_{j_n} g(x)
\label{e15}
\end{eqnarray}
one can easily prove the following relation
\begin{eqnarray}
[ f(x) \, ,^{\!\!\star} g(x) ] &=& 
\sum_{n=1}^\infty{1 \over n!} \left ( {i \over 2} \right )^{n}
 \theta^{i_1 j_1} \dots
\theta^{i_{n} j_{n}}\nonumber\\
&&~~~~\times (1-(-1)^n)  \partial_{i_1} \dots \partial_{i_{n}} f(x) 
                 \partial_{j_1} \dots \partial_{j_{n}} g(x)
\label{e16}
\end{eqnarray}
where $[f,g]$ is an anti-commutator if both $f$ and $g$ are fermionic
 and a commutator otherwise.

The above expression can be written as a total derivative
\begin{eqnarray}
[ f(x) \, ,^{\!\!\star} g(x) ] = 
\partial_i j^i 
\label{e17}
\end{eqnarray}
for a suitable $j^i$, thanks to eq.(\ref{e16}) and the
 anti-symmetry of $\theta$.

Therefore, eq.(\ref{e11}) can be recast in the form
\begin{eqnarray}
{i \over 2} \staranticomm{C}{C} = \partial_i \omega^i 
\label{e18}
\end{eqnarray}
where $\omega^i$ has ghost number two.
By using the above equation eq.(\ref{e11}) reads
\begin{eqnarray}
s \Lambda = \partial_i \omega^i \, ,
\label{e19}
\end{eqnarray}
implying that $\Lambda$ can be characterized as an element of $H(s|d)$,
 the cohomology of $s$ modulo $d$, in the space of the formal
 power series in $a_j$, $c$ and $\theta$. 
Notice that this is an all-order statement with respect to the expansion in
 powers of $\theta$.

The requirement of linearity of $\Lambda$ (and of $C$) in $c$ is 
 guaranteed since $\Lambda$ is of ghost number one and $c$ is the only
 field with positive ghost number and there are no fields with negative
 ghost number.

In summary, the most general solution of eq.(\ref{e6.1}) can be written as
\begin{eqnarray}
C = c + \Xi + \xi \, , ~~~~~ \Xi \in H(s|d) \, , ~~~~~ \xi \in H(s) 
\label{e20}
\end{eqnarray}
where the first term corresponds to the initial condition of 
 eq.(\ref{e4:1}), $\Xi$ is a representative of the cohomology
 of $s$ modulo $d$ and $\xi$ is a representative of the local cohomology of
 $s$.

In the following, we shall check that the coefficients of the solutions for 
 $C^{(1)}$ and $C^{(2)}$ given in eqs.(\ref{e9}) belong in fact to $H(s|d)$.
In particular, we will prove that $C^{(2)}$ is nothing but the Chern-Simons
 three form, up to a total derivative. 

\subsection{Analysis of the coefficients $C^{(1)}$ and $C^{(2)}$}

In this subsection we verify that the coefficients
 $C^{(1)}$ and $C^{(2)}$ in eqs.(\ref{e9}) belong to $H(s|d)$.

Let us consider first $C^{(1)}$. 
Its variation under the BRST operator $s$ gives
\begin{eqnarray}
s C^{(1)} = {1 \over 2} \theta^{ij} \partial_j c \partial_i c =
\partial_j \left ( {1 \over 2} \theta^{ij} c \partial_i c \right ) \, .
\label{e21}
\end{eqnarray}
The latter equality follows from the anti-symmetry of $\theta^{ij}$.
By direct inspection one concludes that $C^{(1)}$ cannot be
 written as a trivial term in the sense of the cohomology of $s$ modulo $d$. 
Thus $C^{(1)}$ identifies a representative of $H(s|d)$.

Let us now look at $C^{(2)}$.
In this case it is not difficult to verify that the expression~(\ref{e9:b})
 can be rewritten as a Chern-Simons term plus a total derivative:
\begin{eqnarray}
C^{(2)} & = & {1 \over 6} \partial_k ( \theta^{kl} \theta^{ij} 
a_l a_j \partial_i c ) \nonumber \\
& & - {1 \over 12} \theta^{kl}\theta^{ij} (a_j f_{kl}
+a_k f_{lj} + a_l f_{jk}) \partial_i c \, .
\label{e22}
\end{eqnarray}
The Chern-Simons term is put in evidence in the last line. 
Let us now check that $C^{(2)}$ belongs to $H(s|d)$.
For that purpose it is sufficient to look at the Chern-Simons. Thus
\begin{eqnarray}
s \left ( \theta^{kl}\theta^{ij} ( a_j f_{kl}
				 + a_k f_{lj}
                                 + a_l f_{jk}) \partial_i c \right ) & = &
\theta^{ij} \theta^{kl} \left ( \partial_j c f_{kl}
                              + \partial_k c f_{lj}
                              + \partial_l c f_{jk} \right ) \nonumber \\
& = & \theta^{ij} \theta^{kl} \left ( \partial_j c f_{kl}
                                    + \partial_k c f_{lj}
                                    + \partial_l c f_{jk} \right . \nonumber \\
& & \left . +c (\partial_j f_{kl} + \partial_k f_{lj} + \partial_l f_{jk} 
) \right ) \partial_i c \, ,
\label{e23}
\end{eqnarray}
where the term in the last line is zero due to the Bianchi identity.
Therefore
\begin{eqnarray}
s \left ( \theta^{kl}\theta^{ij} (a_j f_{kl}
+a_k f_{lj} + a_l f_{jk}) \partial_i c \right ) & = &
\partial_j (\theta^{ij} \theta^{kl} c f_{kl} \partial_i c) \nonumber \\
& +& 
\partial_k (\theta^{ij} \theta^{kl} c f_{lj} \partial_i c) +
\partial_l (\theta^{ij} \theta^{kl} c f_{jk} \partial_i c)\\
&  -& \theta^{ij}\theta^{kl}\left(
	 c f_{kl} \partial_j \partial_i c 
	+ c f_{lj} \partial_k \partial_i c 
    	+ c f_{jk} \partial_l \partial_i c\right) \, . \nonumber 
\label{e25}
\end{eqnarray}
Upon reshuffling of the indices one can see that the terms in the last
 line of the above equation sum up to zero due to the anti-symmetry of the
 $\theta$.
Finally
\begin{eqnarray}
&& s \left ( \theta^{kl}\theta^{ij} (a_j f_{kl}
+a_k f_{lj} + a_l f_{jk}) \partial_i c \right )  = \nonumber \\
&& ~~~~~~~~~~~~~~~~~~~ 
\partial_j (\theta^{ij} \theta^{kl} c f_{kl} \partial_i c) +
\partial_k (\theta^{ij} \theta^{kl} c f_{lj} \partial_i c) +
\partial_l (\theta^{ij} \theta^{kl} c f_{jk} \partial_i c) \, .
\label{e26}
\end{eqnarray}
Thus the BRST variation of $C^{(2)}$ is a total derivative.
Moreover, the presence of the Chern-Simons term ensures that $C^{(2)}$ is
a non-trivial element of the cohomology of $s$ modulo
 $d$~\cite{Barnich:2000zw}.
Its appearance is a consequence of the general results established in the
 previous section.

\subsection{The gauge field $A_i$}

The analysis of the non-commutative Abelian gauge field $A_i$ can be
 performed along the same lines of $C$.
Of course, a solution of
\begin{eqnarray}
s A_i = \partial_i C + i \starcomm{C}{A_i}
\label{e27}
\end{eqnarray}
exists if and only if there is a suitable $\Omega_i$ with ghost number 
 zero such that
\begin{eqnarray}
\partial_i C + i \starcomm{C}{A_i} = s \Omega_i \, .
\label{e28}
\end{eqnarray}
$\Omega_i$ cannot depend on $c$ since it has ghost number zero.
If eq.(\ref{e28}) is verified, from eq.(\ref{e27}) we get
\begin{eqnarray}
s ( A_i - \Omega_i) = 0 \, .
\label{e29}
\end{eqnarray}
Therefore 
\begin{eqnarray}
A_i = \Omega_i + \sigma_i \, , ~~~~~\sigma_i \in H(s) 
\label{e30}
\end{eqnarray}
where $\sigma_i$ is an element of the local cohomology of $s$ in the space
of the formal power series in $a_j$ and $\theta$ with a Lorentz vector
index.

Moreover from eq.(\ref{e17}) we conclude that
\begin{eqnarray}
i \starcomm{C}{A_i} = \partial_k \Omega^k_i 
\label{e31}
\end{eqnarray}
for a suitable $\Omega^k_i$. From eq.(\ref{e28}) we
get
\begin{eqnarray}
s \Omega_i = \partial_k (\Omega^k_i + \delta^k_i C) \, ,
\label{e32}
\end{eqnarray}
which means that also $\Omega_i$ can be characterized in terms of the
 cohomology of $s$ modulo $d$.

The most general solution for the non-commutative gauge field
 $A_i$ can thus be written as
\begin{eqnarray}
A_i = a_i + \Sigma_i + \sigma_i \, ,
 ~~~\Sigma_i\in H(s|d)\,,~~~~\sigma_i\in H(s)
\label{e33}
\end{eqnarray}
where $\Sigma_i$ is an element of $H(s|d)$.
\\
In order to check that the coefficients of the expansion of $A_i$
 in eq.(\ref{e7}) can be identified with elements of $H(s|d)$ let us
 discuss $A^{(2)}_i$, whose expression has been found 
 in~\cite{Jurco:2001rq}.
Up to the first order in $\theta$ it reads
\begin{eqnarray}
A^{(2)}_i = - {1 \over 2} \theta^{kl} a_k (\partial_l a_i + f_{li}) \, .
\label{e34}
\end{eqnarray}
It is interesting to observe that also in this case it is possible to
 put in evidence the Chern-Simons term.
Indeed, up to a total derivative, expression (\ref{e34}) can be written as
\begin{eqnarray}
A^{(2)}_i & = & -{1 \over 2} \partial_l (\theta^{kl} a_k a_i)
+ {1 \over 4} \theta^{kl} a_i f_{lk} - {1 \over 2} \theta^{kl} a_k f_{li} \, .
\label{e35}
\end{eqnarray}
Anti-symmetry of $\theta^{kl}$ and of $f_{li}$ allows us to cast the last
 two terms in the form of a Chern-Simons, namely
\begin{eqnarray}
A^{(2)}_i & = & -{1 \over 2} \partial_l (\theta^{kl} a_k a_i)
+ {1 \over 4} \theta^{kl} ( a_i f_{lk}
+ a_l f_{ki} + a_k f_{il} ) \, ,
\label{e36}
\end{eqnarray}
ensuring that $A^{(2)}_i$ is a non-trivial element of $H(s|d)$.
In much the same way one can easily prove that the BRST variation
 of the coefficient $A_i^{(3)}$ found in~\cite{Okuyama:2001sw} yields
 a total derivative and belongs to $H(s|d)$.

Concerning the matter fields, their inclusion can
 be discussed along the same line. For instance, the first coefficients
 for the non-commutative matter fields can be found in~\cite{Jurco:2001rq},
 and can be easily proven to be characterized in terms of $H(s|d)$ and $H(s)$.

\section{Non-commutative Wess-Zumino consistency condition} \label{sec:WZ}

Although in the Abelian case the expansion in powers of $a_i$ proves
 to be very useful in discussing the cohomological properties of the
 coefficients of $C$ and $A_i$, the expansion in powers of $\theta$
 is better suited to discuss the existence of the SW map.

In this section we provide a complete proof of the existence
 of the SW map in the Abelian case for solutions
 that can be Taylor-expanded in  powers of $\theta$.
The proof will be constructed recursively in the $\theta$ expansion.
The main idea of the proof has been given in~\cite{stora},
 the key ingredient being the associativity of the $\star$-product.

We will focus on the proof of the existence of $C$. The non-commutative
 gauge field $A_i$ can be treated  similarly.
In the course of the proof the nilpotent operator $\Delta$ introduced 
 in~\cite{Brace:2001fj} will naturally arise. 

We wish to prove that the equation
\begin{eqnarray}
s C = i C \star C
\label{e37}
\end{eqnarray}
can be solved by a formal power series
\begin{eqnarray}
C = \sum_{n=0}^\infty C^n(a_j,c,\theta)
\label{e38}
\end{eqnarray}
with the condition $C^0 = c$. $C^n$ is now a local polynomial 
 in $(a_j,c,\theta)$ of order $n$ in $\theta$ and is linear in $c$.

Also, it will be useful to introduce the following notation.
For arbitrary local functions $f(x),g(x)$, we expand the product $f \star g$
 in powers of $\theta$ and define
\begin{eqnarray}
f(x) \star g(x) \equiv \sum_{q=0}^{\infty} f(x) \stargr{q}  g(x) \, .
\label{e39}
\end{eqnarray}
where $ f(x) \stargr{q}  g(x)$ denotes the term of order $q$ in 
the $\theta$ expansion of the $\star$-product.

The proof of the existence will be given by induction in the powers of 
 $\theta$.
At the order zero in $\theta$ eq.(\ref{e37}) is verified since $C^0=c$.
Let us now assume that eq.(\ref{e37}) is verified up to order $n-1$,
 so that 
\begin{eqnarray}
s C^m = i \sum_{p+q+r=m} C^p \stargr{q} C^r \, , ~~~~ m = 0,\dots,n-1 \, .
\label{e40}
\end{eqnarray}
We have to prove that eq.(\ref{e37}) can be verified at the order $n$
 by a suitable choice of $C^n$. At the $n$-th order we get
\begin{eqnarray}
s C^n & = & i \sum_{p+q+r=n} C^p \stargr{q} C^r \nonumber \\
      & = & i C^0 C^n + i C^n C^0 
            + i \mathop{\sum_{p+q+r=n}}_{p \not =n, \, r \not = n}
            C^p \stargr{q} C^r \, ,
\label{e41}
\end{eqnarray}
which can be rewritten as
\begin{eqnarray}
s C^n - i\{ c , C^n \} & = & Z^n \nonumber \\
& \equiv &  i \mathop{\sum_{p+q+r=n}}_{p \not =n, \, r \not = n }
            C^p \stargr{q} C^r \, .
\label{e42}
\end{eqnarray}
Following~\cite{Brace:2001fj} it is useful to introduce the nilpotent 
 operator $\Delta$ defined as
\begin{eqnarray}
\Delta(X) = sX - i [ c, X \} \, ,~~~~\Delta^2=0\,,
\label{e43}
\end{eqnarray}
where $[ c, X \}$ is a commutator if $X$ is bosonic and
 an anti-commutator if $X$ is fermionic.
Eq.(\ref{e42}) takes now the form
\begin{eqnarray}
\Delta C^n = Z^n \, .
\label{e44}
\end{eqnarray}
If a solution $C^n$ of eq.(\ref{e43}) exists, then $Z^n$ must satisfy
 the Wess-Zumino consistency condition
\begin{eqnarray}
\Delta Z^n = 0\, ,
\label{e45}
\end{eqnarray}
due to the nilpotency of $\Delta$.
We will show that eq.(\ref{e45}) holds true due to the associativity
 of the $\star$-product and the recursive assumption in eq.(\ref{e40}).

Some remarks are here in order. In the Abelian case the first two terms in 
 eq.(\ref{e41}) cancel out and the operator $\Delta$ reduces to the usual
 BRST operator $s$.
This is no more true in the non-Abelian case.
However, since the proof of the Wess-Zumino consistency condition 
 in eq.(\ref{e45}) can be carried out along the same lines for both
 the Abelian and the non-Abelian case, we choose to work with the full
 operator $\Delta$.

Let us compute $\Delta Z^n$. We get
\begin{eqnarray}
\Delta Z^n & = & sZ^n - i [c,Z^n] \nonumber \\
           & = & i \mathop{\sum_{p+q+r=n}}_{p\neq n, \, r \neq n}
                 \left(s C^p\right) \stargr{q} C^r 
               - i \mathop{\sum_{p+q+r=n}}_{p\neq n, \, r \neq n}
                 C^p \stargr{q} \left(s C^r\right) \nonumber \\
	   &   & + \mathop{\sum_{p+q+r=n}}_{p \neq n \, , r \neq n}
                 c\ C^p \stargr{q} C^r 
                 - \mathop{\sum_{p+q+r=n}}_{p \neq n \, , r \neq n}
                 C^p \stargr{q} C^r c \, .
\label{e46}
\end{eqnarray}
Taking into account eq.(\ref{e40}), we can rewrite
the BRST variations in the above equation in terms of $\star$-product
 as follows:
\begin{eqnarray}
\Delta Z^n & = &
- \mathop{\sum_{p+q+r+s+t=n}}_{p+q+r \neq n, \, t \neq n}
\left ( C^p \stargr{q} C^r \right ) \stargr{s} C^t \nonumber \\
& & + \mathop{\sum_{p+q+r+s+t=n}}_{p \neq n , \, r+s+t \neq n}
      C^p \stargr{q} \left ( C^r \stargr{s} C^t \right ) \nonumber \\
& & + \mathop{\sum_{p+q+r=n}}_{p \neq n, \, r \neq n} c\ C^p \stargr{q} C^r
    - \mathop{\sum_{p+q+r=n}}_{p \neq n, \, r \neq n} C^p \stargr{q} C^r c \, .
\label{e47}
\end{eqnarray}
We now recast each term appearing in the r.h.s. of the above
 equation in a form convenient to put in evidence 
 the cancellations occurring among the various terms:
\begin{eqnarray}
- \mathop{\sum_{p+q+r+s+t=n}}_{p+q+r \neq n, \, t \neq n}
\left ( C^p \stargr{q} C^r \right ) \stargr{s} C^t & = &
- \sum_{p+q+r+s+t=n} \left ( C^p \stargr{q} C^r \right ) \stargr{s} C^t
\nonumber \\
&& + \sum_{p+q+r=n} \left ( C^p \stargr{q} C^r \right )\ c + c\ c\ C^n \, ,
\nonumber \\
\mathop{\sum_{p+q+r+s+t=n}}_{p \neq n , \, r+s+t \neq n}
    C^p \stargr{q} \left ( C^r \stargr{s} C^t \right ) & = & 
\sum_{p+q+r+s+t=n} C^p \stargr{q} \left ( C^r \stargr{s} C^t \right )
\nonumber \\
&& - c\sum_{p+q+r=n} C^p \stargr{q} C^r - C^n c\ c \nonumber \, , \\
\mathop{\sum_{p+q+r=n}}_{p \neq n, \, r \neq n} c\ C^p \stargr{q} C^r
& = & \sum_{p+q+r=n} c\ C^p \stargr{q} C^r - c\ C^nc - c\ c\ C^n \, ,
\nonumber \\
- \mathop{\sum_{p+q+r=n}}_{p \neq n, \, r \neq n} C^p \stargr{q} C^r c
& = & - \sum_{p+q+r=n} C^p \stargr{q} C^r c + c\ C^n c + C^n c\ c \, .
\label{e48}
\end{eqnarray} 
Finally, summing up all the terms we are left with
\begin{eqnarray}
\Delta Z^{(n)} & = & 
- \sum_{p+q+r+s+t=n} \left ( C^p \stargr{q} C^r \right ) \stargr{s} C^t
+ \sum_{p+q+r+s+t=n} C^p \stargr{q} \left ( C^r \stargr{s} C^t \right ) 
\nonumber \\
& = & 0 \,,
\label{e49}
\end{eqnarray}
where the associativity of the $\star$-product has been used.
Therefore $Z^n$ obeys the Wess-Zumino consistency condition eq.(\ref{e45})
 thanks to the associativity of the $\star$-product and to
 the recursive assumption in eq.(\ref{e40}).

We stress that this result holds for both Abelian and non-Abelian
 gauge groups. 

To conclude that eq.(\ref{e44}) can be solved it remains to
 prove that the cohomology of the operator
 $\Delta$ is empty in the sector with ghost number two.
In particular, if one is able to find out a homotopy operator $K$ for
 $\Delta$
\begin{eqnarray}
\{ K , \Delta \} = {\cal I} \, ,
\label{e50}
\end{eqnarray}
where ${\cal I}$ in the r.h.s. of eq.(\ref{e50}) stands for the identity 
 operator when acting on formal power series depending on $c$
 and zero otherwise,
 then an explicit solution for $C^n$ is provided by
\begin{eqnarray}
C^n =  \Delta K Z^n \, .
\label{e51}
\end{eqnarray}
Eq.(\ref{e51}) follows from the relation
\begin{eqnarray}
Z^n = {\cal I} Z^n = \{ K, \Delta \} Z^n = \Delta K Z^n 
\label{e51.bis}
\end{eqnarray}
which holds true due to the Wess-Zumino
 consistency condition in eq.(\ref{e45}).

For the non-Abelian case such a homotopy operator $K$ has been proposed
 in~\cite{Brace:2001fj}.
For the Abelian case the homotopy operator takes a very simple
 form~\cite{Zumino:Ed.ew}
\begin{eqnarray}
K = \int_0^1 dt \, \left ( 
a_i \lambda_t {\partial \over \partial (\partial_i c)} + 
\partial_{(i} a_{j)} \lambda_t 
{\partial \over \partial (\partial_i \partial_j c)} +
\partial_{(i}\partial_{j} a_{k)} \lambda_t {\partial \over \partial (\partial_i \partial_j \partial_k c)} + \dots 
\right )
\label{e52}
\end{eqnarray}
where $(i,\dots,k)$ denotes total symmetrization and the operator
 $\lambda_t$ acts as
\begin{eqnarray}
&&\lambda_t 
X(c,f_{ij},\partial^n f_{ij},\partial^n c, a_i,
 \partial_{i_1}\dots\partial_{i_{n-2}}\partial_{(i_{n-1}} a_{i_n)}) 
\nonumber\\
&&~~~~~~~~~~~~~~~~= X(c,f_{ij},\partial^n f_{ij},t\partial^n c, t a_i,
 t  \partial_{i_1}\dots\partial_{i_{n-2}}\partial_{(i_{n-1}} a_{i_n)}) \, ,
\label{e52.bis}
\end{eqnarray}
where $X$ is an arbitrary local formal power series.
The operator $\lambda_t$ leaves unchanged the undifferentiated ghost $c$ and
 the field strength $f_{ij}$ and its derivatives $\partial^n f_{ij}$.
The homotopy $K$ allows us to obtain representatives of the higher
 order coefficients in a systematic way.
In the case of the Abelian Freedman-Townsend model the analogue of the
 homotopy operator in eq.(\ref{e52}) can be found in~\cite{Barnich:2001mc}.
The proof outlined for the case of $C$ can be paraphrased  
 along the same lines for the case of $A_i$.

\section{Conclusions}
 
In this work we have provided a cohomological interpretation of the solution
 of the Seiberg-Witten (SW) map for Abelian gauge theories in terms
 of $H(s|d)$, the cohomology of the Abelian BRST differential $s$
 modulo $d$, with the addition of elements of $H(s)$, the local cohomology
 of $s$.
This feature might help in clarifying the geometrical and topological nature
 of the SW map.
The examples of the first coefficients for the non-commutative fields
 $A_i$ and $C$ have been analysed, and shown to be written in terms 
 of the Chern-Simons three form.

It is worth mentioning here that, due to the presence of the $\theta$'s,
 suitable generalizations of the Chern-Simons are to be expected for the
 higher order coefficients.
Consider for instance the term of order three in $a_i$
\begin{eqnarray}
\theta^{km}\theta^{pq}\left(
a_{[i} f_{km]} f_{pq}+
a_{[p} f_{km]} f_{qi}+
a_{[q} f_{km]} f_{ip}\right)\label{CS5}
\end{eqnarray}
where $[ikm]$ means complete antisymmetrization in the Lorentz indices
 $i$, $k$, $m$.
By using the Bianchi identity, it is almost immediate to verify that the BRST
 variation of the term~(\ref{CS5}) gives a total derivative.
On the other hand expression~(\ref{CS5}) turns out to be BRST non trivial.
We see thus that the presence of the $\theta$'s allows for
 the existence of new elements of the cohomology of $s$ modulo $d$,
 which would be automatically vanishing if the only antisymmetric
 quantity at our disposal would be the Levi-Civita tensor
 $\epsilon^{ijkl}$.

The full characterization of $H(s|d)$ in the presence of the
 $\theta$'s is therefore of primary importance in order to work out
 the most general solution of the SW map.
This is an important point which deserves further
 investigation~\cite{WorkInProgress}.

We have also discussed the Wess-Zumino consistency condition on general
 grounds, proving that it can be fulfilled order by order in $\theta$
 for both Abelian and non-Abelian case, due to the associativity of the
 $\star$-product.
This condition plays a key role in the proof of the existence of the SW map.
It is indeed a necessary condition.
It becomes a sufficient condition provided one is able to show that the
 cohomology of the coboundary operator $\Delta$ is empty in the sector of
 ghost number two and one, which would ensure the existence of $C$ and $A_i$.
The homotopy operator for the differential $\Delta$
 in the Abelian case has been used in order to find an explicit solution for
 the SW map.
This concludes the proof of the existence of the map in this case.

As a final comment, we emphasize that the cohomological analysis carried out
 here concerns only classical aspects.
Moreover, the knowledge of the most general solution of the SW map should be
 the natural starting point for a possible consistent quantization of the 
 non-commutative theories.
Indeed, there are indications that the intrinsic ambiguities of the SW map
 show up at quantum level even if they are not included in the starting
 classical action~\cite{Jurco:2001rq,Bichl:2001cq}.
 
\section*{Acknowledgements}

A.~Quadri is deeply indebted to Professor R.~Stora for the fundamental help
in the proof of the existence of the Seiberg-Witten map and for many
 enlightening discussions on the subject.
S.P.~Sorella thanks the Theoretical Physics Department of the Milano University
(Italy) and the INFN (sezione di Milano) for the kind invitation.
The Conselho Nacional de Desenvolvimento
Cient\'{\i}fico e Tecnol\'{o}gico CNPq-Brazil, the Funda{\c {c}}{\~{a}}o de
Amparo {\`{a}} Pesquisa do Estado do Rio de Janeiro (Faperj), the SR2-UERJ,
the INFN (sezione di Milano) and the MIUR, Ministero dell'Istruzione
 dell'Universit\`a e della Ricerca, Italy, are acknowledged for the
financial support.

\end{document}